\documentclass[12pt]{article}

\usepackage{amssymb}
\usepackage{amsmath}
\usepackage{latexsym}
\usepackage{yfonts}

\catcode`@=11
\renewcommand{\section}{\@startsection{section}{1}{0pt}{\medskipamount}
{\medskipamount}{\large\bf}} \numberwithin{equation}{section}
\catcode`@=12

\def\beq{\begin{eqnarray}}    
\def\eeq{\end{eqnarray}}      

\def\={\ =\ }

\begin{document}

\begin{center}
{\Large\bf General conversion method for
constrained systems 
}

\vspace{18mm}

{\Large Igor A. Batalin$^{(a,b)}\footnote{E-mail:
batalin@lpi.ru}$\;,
Peter M. Lavrov$^{(b, c)}\footnote{E-mail:
lavrov@tspu.edu.ru}$\;
}

\vspace{8mm}

\noindent ${{}^{(a)}}$
{\em P.N. Lebedev Physics Institute,\\
Leninsky Prospect \ 53, 119 991 Moscow, Russia}

\noindent  ${{}^{(b)}}
${\em
Tomsk State Pedagogical University,\\
Kievskaya St.\ 60, 634061 Tomsk, Russia}

\noindent  ${{}^{(c)}}
${\em
National Research Tomsk State  University,\\
Lenin Av.\ 36, 634050 Tomsk, Russia}

\vspace{20mm}

\begin{abstract}
\noindent We reformulate in a systematic way the conversional
approach in its most general and compact form. We present a new
definition of generalized Dirac bracket directly in terms of the
super-observables commuting with the basic BFV-BRST charge.
\end{abstract}

\end{center}

\vfill

\noindent {\sl Keywords:} Constrained systems, conversion method, Dirac bracket,
BFV-BRST charge
\\

\noindent PACS numbers: 11.10.Ef, 11.15.Bt
\newpage

\section{Introduction}

In the Dirac theory of Hamiltonian constraint dynamics, all constraints
are split naturally into the two classes \cite{Dirac,BGol}.
First-class constraints are in Poisson-bracket involution among themselves.
So, they do serve naturally as a
gauge symmetry generators. Second-class constraints have their
Poisson-bracket matrix invertible. So, they do
reduce effectively an original phase space to the second-class constraint hypersurface.
Locally, in their Abelian
form, first-class constraints do commute among themselves, so
that they are similar, say, to a set of momenta.
Second-class constraints, in their local Abelian form,
have their Poisson-bracket matrix invertible and constant.
So, they are similar to a set of canonical pairs of co-ordinates and conjugate momenta.

The famous Dirac bracket concept provides for a natural projection to the
Poisson bracket  to a tangential subspace
with respect to the second-class constraint hypersurface.
However, it appears  a rather difficult problem as to how
to reformulate the Dirac bracket concept within a consistent quantum theory.
In a series of papers \cite{FadShat,EgM,BF2,BF1,BT,FL1,FL,Fedosov,BGL,BG,BGL1,BL}
the so-called
conversional approach to the quantization of dynamical systems
with second-class constraints has been developed.
This approach is based on the idea of converting
the second-class-constraints into effective first-class ones by introducing
extra degrees of freedom. Initial first-class
constraints (if they are present in the system) and the initial Hamiltonian
should also be converted into the corresponding modified
objects, depending on the extra variables, so that we have,
as a result, a new Hamiltonian in involution with new
constraints of the first class only.

Thus, within the framework of the conversional approach,
the problem of quantizing the system with general constraints
is in fact reduced to the case of first-class constraints only,
for which the scheme of generalized canonical quantization,
which is operating well, does exist \cite{BF2,BF1,BFF1,BF3,BFF3}. Thereby, the unification
proposed does resolve the operator quantization problem,
whereas one has to make use of the canonical commutation relations, only.

In  the present article, we reformulate systematically the
conversional approach in its most general and compact form. We
present a new definition for the Dirac bracket directly in terms of
the super-observables commuting with the basic BFV-BRST charge.

NOTATIONS: $\{ A, B \}$ and $[A, B ]$ denotes the Poisson (super)bracket and the
(super)commutator, respectively. $\varepsilon(A)$ and ${\rm gh(A)}$ denotes the
Grassmann parity and the ghost number, respectively. Other notation is
clear from the context.

\section{Conversion of constraints in its most general form}

Let
\beq
\label{CC1.1}
Z  =:  ( p, q ); \quad  \varepsilon( p ) = \varepsilon( q ),\quad
{\rm gh}( p )  = -  {\rm gh}(
q )  =:  0   
\eeq
be a set of initial canonical variables, and let
\beq
\label{CC1.2}
\phi^{ \alpha };\quad  \varepsilon( \phi^{ \alpha } ) =: \varepsilon_{ \alpha },\quad
{\rm gh}( \phi^{ \alpha } )  =  0,     
\eeq
be the conversion variables  commuting as
\beq
\label{CC1.3}
\{ \phi^{ \alpha },  \phi^{ \beta } \}  =:  \omega^{ \alpha\beta }  =  {\rm const},
\eeq
with even invertible metric $\omega^{ \alpha\beta }$,
\beq
\label{CC1.4}
\varepsilon( \omega^{ \alpha\beta } )  =  \varepsilon_{ \alpha }  +
\varepsilon_{ \beta }.  
\eeq
In turn, let
\beq
\label{CC1.5} C^{A},  \mathcal{ P }_{B};\quad
 \varepsilon( C^{A} ) = \varepsilon( \mathcal{
P}_{A} )  =: \varepsilon_{A} +
1,  \quad   
{\rm gh}( C^{ A } ) =  - {\rm gh}( \mathcal{ P }_{ A } )  =:  1,      
\eeq
be the ghost canonical variables
\beq
\label{CC1.8}
\{ C^{ A } , \mathcal{ P }_{ B }  \}  =:  \delta^{ A }_{ B },    
\eeq

Define the "BFV - BRST" charge,
\beq
\label{CC1.9}
Q  =:  Q( Z, \phi, C, \mathcal{ P } ), \quad  \varepsilon( Q )  =:  1,   \quad
{\rm gh}( Q) =: 1,  
\eeq
to satisfy the master equation,
\beq
\label{CC1.10}
\{ Q, Q \}  =  0,     
\eeq
and the boundary condition
\beq
\label{CC1.11}
Q  = C^{ A } \mathcal{ T }_{ A }( Z, \phi )  +  . . .    ,    
\eeq
where ellipses mean higher powers in ghosts (\ref{CC1.5}).

If one expands the $Q$ to the first order in ghost momenta $\mathcal{ P }$,
\beq
\label{CC1.12}
Q  =  C^{ A } \mathcal{ T }_{ A }( Z, \phi )  +
  \frac{ 1 }{ 2 }  (-1)^{ \varepsilon_{ B } }  C^{ B } C^{ A }
\mathcal{ U }_{ AB }^{\; C } ( Z, \phi ) \mathcal{ P }_{ C } (-1)^{
\varepsilon_{ C } }  +  . . .     ,     
\eeq
then the involution relations follow from the master equation (\ref{CC1.10}),
\beq
\label{CC1.13}
\{ \mathcal{ T }_{ A },  \mathcal{ T }_{ B } \}  =  \mathcal{ U }_{ AB }^{ \;C }
\mathcal{ T }_{ C }.    
\eeq
These relations show us that the coefficients $\mathcal{ T }_{ A }( Z, \phi )$
are effective
(converted) first-class constraints in the "extended original phase space"
spanned with the phase variables ( $Z, \phi$ ). These effective first-class constraints
can be split as
\beq
\label{CC1.14}
\mathcal{ T }_{ A }  =  (  T_{ a }( Z, \phi ); \;\! \Theta_{ \alpha}( Z, \phi
)  ),   
\eeq
where  $\varepsilon( \Theta_{ \alpha } )  =:  \varepsilon_{ \alpha }$, the
second in (\ref{CC1.2}),  with
\beq
\label{CC1.15}
T_{ a }( Z, 0 )  =:  t_{ a }( Z ) \;\;{\rm and}\;\;  \Theta_{ \alpha }( Z, 0 )  =:
\theta_{ \alpha }( Z )      
\eeq
being  original first-class and second-class constraints, respectively.

Define an observable  $A(  Z, \phi, C, \mathcal{ P} )$
as to satisfy  the standard equation,
\beq
\label{CC1.16}
\{ Q,  A \}  =  0.    
\eeq
For two observables, $A$ and $B$, the generalized Dirac bracket,
$\{  \:,  \:\}_{ D }$, is then defined as
\beq
\label{CC1.17}
\{ A_{ 0 }, B_{ 0 } \}_{ D }  =:  \{ A, B \}_{ 0 },   
\eeq
where it is denoted for an arbitrary $X$,
\beq
\label{CC1.18}
X_{ 0 }  =:  X |_{ \phi  =  0 }.      
\eeq
By expanding the $A, B$ and $Q$ in power series in $\phi$,
\beq
\label{CC1.19}
&&A  =  A_{0}  +  \phi^{ \alpha }  A_{ \alpha}  +   . . .  ,  \quad
   B  =  B_{0}  +  \phi^{ \alpha }  B_{ \alpha }  +  . . .  , \\   
\label{CC1.20}
&&Q  =  Q_{0}  +  \phi^{ \alpha } \;\! Q_{ \alpha }   +   . . .  ,   
\eeq
we rewrite (\ref{CC1.17}) as
\beq
\label{CC1.21}
\{ A_{0}, B_{0} \}_{ D }  =  \{ A_{0}, B_{0} \}  +
A_{ \alpha } \;\! \omega^{ \alpha\beta }
B_{ \beta }  (-1)^{  ( \varepsilon(A) + 1 )  \varepsilon_{ \alpha } }, 
\eeq
where $A_{ \alpha }, B_{ \alpha }$ and $Q_{ \alpha}$  should satisfy the equations
\beq
\label{CC1.22}
&&Q_{ \alpha }\;\! \omega^{ \alpha\beta } A_{ \beta }  =  -  \{ Q_{0}, A_{0} \},  \\
\label{CC1.23}
&&Q_{ \alpha }\;\! \omega^{ \alpha\beta } B_{ \beta }  =  -  \{ Q_{0}, B_{0 } \},  \\
\label{CC1.24}
&&\{ Q_{0}, Q_{0} \}  +  Q_{ \alpha }\;\! \omega^{ \alpha\beta } Q_{ \beta }  =  0. 
\eeq
These equations rewrite themselves  in a natural way in terms of the definition (\ref{CC1.21})
\beq
\label{CC1.24a}
\{Q_0,A_0\}_{D}=0,\quad \{Q_0,B_0\}_{D}=0,\quad \{Q_0,Q_0\}_{D}=0 .
\eeq

Due to the ghost number conservation, these equations are uniquely resolvable,
 certainly.
Indeed, in the Abelian second-class constraint basis, we have
\beq
\label{CC1.25}
&&A_{ \alpha }  = - \{ \Upsilon_{ \alpha }( Z ),  A_{0}( Z ) \}
(-1)^{ \varepsilon_{ \alpha} },  \\
\label{CC1.26}
&&B_{ \alpha }  = - \{ \Upsilon_{ \alpha }( Z ),  B_{0}( Z ) \}
(-1)^{ \varepsilon_{ \alpha } }, \\  
\label{CC127} &&\{  \Upsilon_{ \alpha },  \Upsilon_{ \beta }  \}  =
\omega_{ \alpha\beta }(-1)^{ \varepsilon_{\alpha} },
 \eeq
 and
 \beq
\label{CC1.28} &&Q_{0}  =  C^{ \alpha } \Upsilon_{ \alpha }( Z )  +
{\rm terms \;\;independent \;\;of}\;\; C^{ \alpha },   \\
\label{CC1.29}
&&Q_{ \alpha }  = -\; \omega_{ \alpha\beta }\;\! C^{ \beta }  +
{\rm terms\;\; independent\;\; of}\;\; C^{ \alpha }.  
\eeq

In order to cover the case of the general basis of second-class constraints
$\theta_{\alpha}( Z )$, we define an even  matrix
\beq
\label{CC1.30}
V_{\alpha}^{\;\beta}( Z ), \quad
\varepsilon( V_{\alpha}^{\;\beta} ) = \varepsilon_{\alpha} + \varepsilon_{\beta}, 
\eeq
so as to satisfy the equation
\beq
\label{CC1.32}
\{ \theta_{\alpha}, \theta_{\beta} \}  =
V_{\alpha}^{\;\gamma}  (-1)^{ \varepsilon_{\gamma} }  \omega_{\gamma \delta}
V_{\beta}^{\;\delta}  (-1)^{ ( \varepsilon_{\beta} + 1 ) \varepsilon_{\delta} } . 
\eeq
In terms of the latter matrix (\ref{CC1.30}), the equations
(\ref{CC1.25})-(\ref{CC1.29}) modify as
\beq
\label{CC1.33}
&&\{ \theta_{\alpha}, A_{0} \} =
- V_{\alpha}^{\;\beta}  (-1)^{ \varepsilon_{\beta} }  A_{\beta}, \\  
\label{CC1.34}
&&\{ \theta_{\alpha}, B_{0} \} =
- V_{\alpha}^{\;\beta}  (-1)^{ \varepsilon_{\beta} }  B_{\beta},  \\
\label{CC1.35}
&&\{ \theta_{\alpha}, \theta_{\beta} \}  D^{\beta \gamma}  =
\delta_{\alpha}^{\; \gamma},  \\ 
\label{CC1.36}
&&Q_{0} = C^{\alpha} \theta_{\alpha}( Z )  + . . . \;     ,  \\
\label{CC1.37}
&&Q_{\gamma} = C^{\alpha}  V_{\alpha}^{\;\beta}
(-1)^{ \varepsilon_{ \beta} }  \omega_{\beta \gamma}  + . . .  \;   .   
\eeq
It follows  then the standard formula for the Dirac bracket
(\ref{CC1.21}),
\beq
\label{CC1.38}
\{ A_{0}, B_{0} \}_{D} = \{
A_{0}, B_{0} \} - \{ A_{0}, \theta_{\alpha} \}  D^{\alpha \beta}  \{
\theta_{\beta}, B_{0} \}. \eeq
Here in (\ref{CC1.33}),
(\ref{CC1.34}), (\ref{CC1.38}), we have assigned zero  values as for
all ghost variables, which means  the lowest order in ghosts. Also,
here we do assume, for the sake of simplicity, that the lowest
structure coefficients ${\cal U}_{ \alpha \beta }^{\gamma}$ and ${\cal U}_{ \alpha
\beta }^{c}$ in (\ref{CC1.12}) are zero at $\phi^{\alpha} = 0$
(Abelian conversion).

It follows directly from (\ref{CC1.32})-(\ref{CC1.35}) that
\beq
\label{2.36}
-  \{ A_{0}, \theta_{\alpha} \} D^{\alpha \beta} \{ \theta_{\beta},
B_{0} \} = A_{\delta} \tilde{V}_{\;\;\alpha}^{\delta} (-1)^{ ( \varepsilon_{A} +
1) \varepsilon_{\delta} }( \tilde{V}^{-1} )_{\;\;\mu}^{\alpha}\; \omega^{\mu\nu}
(-1)^{\varepsilon_{\nu}} ( V^{-1} )_{\nu}^{\;\;\beta}
V_{\beta}^{\;\;\gamma} (-1)^{\varepsilon_{\gamma}} B_{\gamma},
\eeq
where
\beq
\label{2.37}
\tilde{V}_{\;\;\mu}^{\alpha} =: V_{\mu}^{\;\alpha} (-1)^{
\varepsilon_{\alpha} (\varepsilon_{\mu} + 1) },  
\eeq
is a super-transposed to $V$. Now, the $V$ drops out completely from
(\ref{2.36}), and we arrive at (\ref{CC1.38}).

If the coefficients $\mathcal{U}_{\alpha\beta}^{\gamma}$
and/or $\mathcal{U}_{\alpha\beta}^{c}$
are non-zero at $\phi^{\alpha} = 0$, then one should shift in
(\ref{CC1.32}), (\ref{CC1.35}):
\beq
\label{2.38}
\{\theta_{\alpha}, \theta_{\beta}\}\;\rightarrow \; \{\theta_{\alpha}, \theta_{\beta}\}  -
\mathcal{U}_{\alpha\beta}^{\gamma} \theta_{\gamma} -
\mathcal{U}_{\alpha\beta}^{c} t_{c} ,
\eeq
which means a symptom of a non-Abelian conversion.

The standard conversion procedure has been considered perturbatively
via $\phi$-power series expansion in Refs.
\cite{FadShat,EgM,BF2,BF1,BT,FL1,FL,Fedosov,BGL,BG,BGL1,BL}, as
applied to the simplest particular cases of linear and Abelian
conversions, and then to the general case of non-Abelian conversion.
The latter allows one to deal with non-scalar constraints, as well.

\section{Operator formulation}

In the previous Section 2, we did consider constraint dynamics at
the classical level, in terms of the canonical Poisson brackets.
Now, we are in a position as to consider in short  how to apply the
Dirac formal quantization rule. First, we change all classical
quantities for respective operators. Then, we change all Poisson
brackets for respective (super) commutators, \beq \label{CC3.1} \{
\; , \; \} \;\rightarrow \; ( i \hbar)^{-1} [ \; , \; ]   ,  \quad
 [ A, B ]  =:  A B  -  B A  (-1)^{ \varepsilon( A ) \varepsilon( B ) }.    
\eeq
In this way, we reformulate our basic master equation (\ref{CC1.10}) as
\beq
\label{CC3.2}
[ Q, Q ]  = 0, \quad
[ C^{A}, \mathcal{ P }_{B} ]  = i \hbar \;\!\delta^{A}_{B} 1, 
\eeq
Further, we consider the (\ref{CC1.12}) as a  $C \mathcal{ P}$
normal ordered power series expansion for the
operator $Q$. Coefficients in (\ref{CC1.12}) are operator valued
functions of the operators (\ref{CC1.1}), (\ref{CC1.2}), now
commuting as
\beq
\label{CC3.3}
&&[ q^{j}, p_{k} ]   =:   i \hbar \;\! \delta^{j}_{k}\;\!  1,\\  
\label{CC3.4}
&&[ \phi ^{ \alpha}, \phi^{ \beta } ]   =:   i \hbar \;\! \omega^{ \alpha \beta } 1.
\eeq Here, we are not interested, so far, as to which type of normal
ordering is chosen for those operators (\ref{CC1.1}), (\ref{CC1.2}).
To the zeroth order in ghost momenta, it follows from  (\ref{CC3.2})
\beq \label{CC3.4} [ \mathcal{ T }_{ A }, \mathcal{ T }_{ B } ]  =
i \hbar \;\!\mathcal{ U }_{ AB }^{\; C }  \mathcal{ T }_{ C },  
\eeq
which looks quite similar to the classical involution (\ref{CC1.13}).
The latter similarity holds because the $C \mathcal{ P }$
normal ordering chosen does respect the ghost numbers of  $C$ and $\mathcal{ P }$.
Consider, however, the Jacobi
relations that follow from (\ref{CC3.2}) to the first order in ghost momenta
$\mathcal{ P }$,
\beq
\nonumber
&&\left(   ( i \hbar )^{-1} [  \mathcal{ U }_{ AB }^{\; E },  \mathcal{ T }_{ C }  ]
(-1)^{ \varepsilon_{ C } \varepsilon_{ E } }  +
 \mathcal{ U }_{ AB }^{\; D }\;\! \mathcal{ U }_{ DC }^{ \;E }   \right)
 (-1)^{ \varepsilon_{ A }\varepsilon_{ C } }  +\\
\label{CC3.5}
 &&+ \;{\rm cyclic \;\;permutations}\; ( A, B, C )   +
  \frac{ 1 }{ 2 }\;\! \mathcal{ U }_{ ABC }^{\; FD }\;\!  \Pi_{ DF }^{ E }  =  0.    
\eeq Here in (\ref{CC3.5}), the operator $\mathcal{ U }_{ABC}^{ \;FD
}$ enters the $CCC\mathcal{ P }\mathcal { P }$ order in
(\ref{CC1.12}), \beq \label{CC3.6} \frac{ 1 }{ 12 }  C^{ C } C^{ B }
C^{ A } ( -1)^{ \varepsilon_{ A } \varepsilon_{ C } + \varepsilon_{
B } } \mathcal{ U }_{ ABC }^{ \;FD }\;\!  \mathcal{ P }_{ D }
\mathcal{ P }_{ F }  (-1)^{ \varepsilon_{ D } },      
\eeq
while the operator
\beq
\label{CC3.8}
\Pi_{ DF }^{ E }  =:  \mathcal{ T }_{ D } \delta_{ F }^{ E }  -
( D \; \leftrightarrow  \;F )  (-1)^{ \varepsilon_{ D } \varepsilon_{ F } }  -
i  \hbar \;\! \mathcal{ U }_{ DF }^{ E },    
\eeq
annihilates the constraint operators,
\beq
\label{CC3.9}
\Pi_{ DF }^{ E }  \mathcal{ T } _{E}  =  0,    
\eeq
due to the (\ref{CC3.4}).  Thereby, we have confirmed the compatibility
of the operator valued involution relations (\ref{CC3.4}).
All higher compatibility relations can be confirmed subsequently
by making use of the generating Jacobi  identity,
\beq
\label{CC3.10}
[ Q, [ Q, Q ] ]  =  0.    
\eeq
Now, we can see from (\ref{CC3.8}) that, in contrast to the involution (\ref{CC3.4}),
the first Jacobi relation (\ref{CC3.5}) has acquired
an actual quantum correction ( the third term in (\ref{CC3.8})),
as compared to the classical counterpart to the (\ref{CC3.5}).
Also, it seems worthy to mention again that, in general,
actual quantum corrections could appear already
in the involution of constrains when using another normal ordering
for ghosts, such as the Weyl or the Wick
ordering.

If one defines the $Q$-invariant converted constraints (they are
similar to the BRST-invariant constraints \cite{BT1} in relativistic
field theory),
\beq
\label{CC3.11}
T_{A} =:  ( i \hbar )^{-1} [
\mathcal{P}_{A}, Q ] (-1)^{ \varepsilon_{A} } ,\quad
[ T_{A}, Q ]  =  0,  
\eeq then their gauge algebra is generated by the relations via the
procedure of \cite{BL1} \beq \label{CC3.12} ( i \hbar )^{-1} [ T_{A}
, T_{B} ]  =  ( i \hbar )^{-3}  [ ( \mathcal{P}_{A}
(-1)^{\varepsilon_{A}},
\mathcal{P}_{B} (-1)^{\varepsilon_{B}} )_{Q}, Q ],  
\eeq where the general quantum antibracket, $( A, B )_{Q}$, is
defined by
\beq
\label{CC3.13} ( A, B )_{Q}  =:  \frac{1}{2} (  [ A,
[ Q, B ] ]  -
( A \;\leftrightarrow \; B ) (-1)^{ ( \varepsilon(A) + 1 ) ( \varepsilon(B) + 1 ) }  ), 
\eeq as for any two operators $A$ and $B$ \cite{BM1,BM2,K-S,Ber}. It follows from
(\ref{CC3.13})
\beq
\label{CC3.14}
[ Q, ( A, B )_{Q} ]  =  [ [Q, A ], [ Q, B ] ].    
\eeq
By choosing in (\ref{CC3.14}) $A = \mathcal{P}_{A}$, $B = \mathcal{P}_{B}$,
one arrives at (\ref{CC3.12}).

\section{Intrinsic Weyl symbols as for conversion variable operators}

Let us proceed from the master equation (\ref{CC1.10}),
to consider its Weyl symbol representation
with respect to the conversion variable operators $\phi^{ \alpha }$ commuting
as in (\ref{CC3.4}). We do
proceed from the Weyl representation as
for any operator $X( Z, \phi, C, \mathcal{ P } )$,
\beq
\label{CC5.1}
X  \; \leftrightarrow \; \tilde{ X }, \quad
      X Y \;\leftrightarrow\;\tilde{ X } \star \tilde{ Y },  
\eeq
with $\tilde{ X }$ being a Weyl symbol as for an operator $X$,
\beq
\label{CC5.2}
X  =:  \exp\left\{  \phi^{ \alpha }
\frac{ \partial }{ \partial \tilde{ \phi }^{ \alpha } }  \right\}
\tilde{ X }(  Z, \tilde{ \phi }  ) |_{ \tilde{ \phi }  =  0 },   
\eeq
where in (\ref{CC5.2}),  $\tilde{ \phi }^{ \alpha }$
means ordinary classical variables.
It follows then from (\ref{CC1.10})
\beq
\label{CC5.3}
\tilde{ Q } \star \tilde{ Q }  =  0.     
\eeq
In  particular, as for the quantum involution (\ref{CC3.4}), it follows
\beq
\label{CC5.4}
\tilde{ \mathcal{ T } }_{ A }  \star  \tilde{ \mathcal{ T } }_{ B }  -
( A \; \leftrightarrow \; B )  (-1)^{ \varepsilon_{ A } \varepsilon_{ B } }  =i\hbar\;\!
\tilde{ \mathcal{ U } }_{ AB }^{\; C }  \star  \tilde{ \mathcal{ T } }_{ C }. 
\eeq
Here, in the second in (\ref{CC5.1}), (\ref{CC5.3}), (\ref{CC5.4}),
the $\star$ means the Weyl symbol multiplication,
\beq
\label{CC5.5}
\star  =:  \exp\left\{  \frac{ i \hbar }{ 2 }
\frac{ \overleftarrow{\partial } }{ \partial \tilde{ \phi }^{ \alpha } }
\omega^{ \alpha\beta }
\frac{ \overrightarrow{\partial } }{ \partial \tilde{ \phi }^{ \beta } }  \right\},  
\eeq
Similarly to (\ref{CC5.4}), the symbol representation can easily
be derived as for the first Jacobi relation
(\ref{CC3.5}), as well as for all higher Jacobi relations.
By using the symbol representations, one can
also expand easily the respective relations
in power series in the classical variables $\tilde{ \phi }$,
as to derive the relations required for their tensor valued coefficient operators.

 In terms of a symbol super-commutator,
\beq
\label{CC5.6}
[ \tilde{A}, \tilde{B} ]_{\star}  =:  \tilde{A}  \star  \tilde{B}  -
\tilde{B}  \star  \tilde{A}
(-1)^{ \varepsilon( \tilde{A} ) \varepsilon( \tilde{B} ) },  
\eeq
one can consider the equations for symbols of physical observables,
$\tilde{A}, \tilde{B}$,
\beq
\label{CC5.7}
[ \tilde{Q}, \tilde{A} ]_{\star}  =  0,\quad
[ \tilde{Q}, \tilde{B} ]_{\star}  =  0,   
\eeq
so as to define the symbol Dirac's bracket,
\beq
\label{CC5.8}
[ \tilde{A}_{0}, \tilde{B}_{0} ]_{D}  =:
( [ \tilde{A}, \tilde{B} ]_{\star} )_{0},   
\eeq
where we have denoted,
\beq
\label{CC5.9}
X_{0}  =:  X|_{ \tilde{ \phi }  =  0 },  \quad  {\rm for \;any\; X}.   
\eeq

\section{Discussion}

It is an important aspect of the conversion method, what is the
relativistic status of the conversion variables.
So far, the latter question remains open in its general meaning.
In principal, if one proceeds from relativistic
covariant Lagrangian theory, it seems natural to expect
the relativistic covariance group to be represented in
the Hamiltonian formalism, in the form of the respective algebra
in terms of Poisson brackets. However, when
converting second-class constraints, one introduces extra conversion variables,
quite new with respect to the
original theory.  So, their  relativistic status expected is also unclear originally.
Moreover, it remains unclear
originally,  which type and form of the effective (converted) gauge algebra we could
expect to be compatible
with required relativistic covariance. Another open question concerns the boundary
condition for converted
constraints. Usually, we do assume the natural boundary conditions requiring
the converted constraints to
coincide with the original second-class constraints at zero value of the conversion variables.
However, it is
unknown if such boundary conditions do respect the relativistic covariance.
Besides, it is worthy to mention
that taking the zero value of the conversion variables is by itself a particular
case of second-class constraints,
although very simple. To avoid that point, when expanding the converted constraints
in power series in the
conversion variables, we identify directly the zeroth order term with the original
second-class constraints.
Of course, we find ourselves rather far from being able to provide for general answers
to even some of the questions mentioned. Our main conjecture is the following.
Being the conversion variables introduced in an appropriate way, so
that they have their relativistic status well-defined, one can expect the relativistic
covariance transformations to be
realized in the form of canonical transformations, typical for all other symmetry
transformations. In particular, we do
not insist on being the natural boundary conditions  the only possibility.
It seems natural to expect that one should
apply some canonical transformation to the "naturally converted" constraints,
as to make them  well-defined in their
relativistic status. Now, we are in a position to try to demonstrate
what we mean by considering a simple example.

Consider first the second-class constraints in the Proca model \cite{Proca},
\beq
\label{CC6.1}
\Theta_{ P }  =:  ( \pi^{ 0 } ;  \pi^{ i }_{, i } + m^{2} A_{ 0 } ),       
\eeq
with $\pi^{ 0}$ and $\pi^{ i }$ being canonical momenta conjugate to $A_{ 0 }$ and $A_{ i }$,
respectively.
The first-class constraints converted from (\ref{CC6.1}) under natural boundary conditions are
\beq
\label{CC6.2}
T_{ P }  =:  \Theta_{ P }  +  m ( \phi ; p ),      
\eeq
with  $\phi$ and $p$ being the conversion field and its conjugate momentum, respectively.
On the other hand, consider the original first-class constraints in the Stueckelberg model
\cite{Stueck},
\beq
\label{CC6.3}
T_{ S }  =:  ( \pi^{ 0 } ; \pi^{ i }_{ , i } + m p ).   
\eeq
Here in (\ref{CC6.3}), we have identified the Stueckelberg scalar field
and the Proca conversion field in (\ref{CC6.1}).
Regrettably, the (\ref{CC6.2}) does not coincide with the (\ref{CC6.3}).
However, it follows immediately that
\beq
\label{CC6.4}
U^{-1} T_{ S } U  =  T_{ P },      
\eeq
with $U$ being a unitary transformation of the form
\beq
\label{CC6.5}
U  =:  \exp\left\{ \frac{ i }{ \hbar } m A_{ 0 } \phi \right \}.   
\eeq

\section*{Acknowledgments}
\noindent The authors  would like  to thank Klaus Bering of Masaryk
University for interesting discussions. The work of I. A. Batalin is
supported in part by the RFBR grant 17-02-00317. The work of P. M.
Lavrov is supported by the Ministry of Education and Science of
Russian Federation, grant  3.1386.2017 and by the RFBR grant
18-02-00153.
\\

\end{document}